\documentclass[twocolumn,english,showpacs]{revtex4}
\usepackage{amsmath,amssymb,dcolumn}
\usepackage{epsfig}
\def\imo{i}
\begin{document}
\title{Instability of higher dimensional charged black holes in the de Sitter world}
\author{R. A. Konoplya}\email{konoplya_roma@yahoo.com}
\affiliation{Department of Physics, Kyoto University, Kyoto 606-8501, Japan}
\author{A. Zhidenko}\email{zhidenko@fma.if.usp.br}
\affiliation{Instituto de F\'{\i}sica, Universidade de S\~{a}o Paulo
C.P. 66318, 05315-970, S\~{a}o Paulo-SP, Brazil}
\begin{abstract}
We have shown that higher dimensional Reissner-Nordstr\"om-de Sitter black holes are gravitationally unstable for large
values of the electric charge and cosmological constant in $D \geq 7$ space-time dimensions.
We have found the shape of the slightly perturbed black hole at the threshold point of instability.
Why only $D=4, 5$ and $6$ dimensional worlds are favorable as to the black stability remains unknown.
\end{abstract}
\maketitle

\textbf{Introduction.}   The issue of stability of black holes was addressed for the first time yet in 1956, in a seminal paper of Regge and Wheeler \cite{Regge:1957td}, who showed that four-dimensional Schwarzschild black holes, are stable against gravitational perturbations. This result confirmed that the Schwarzschild solution can indeed  describe neutral, non-rotating black holes, because gravitationally unstable systems simply could not exist. Later, the stability analysis was generalized for the Reissner-Nordstr\"om and Kerr solutions, which describes the electromagnetically charged non-rotating \cite{RN} and neutral rotating black holes \cite{Kerr}. In 1992, the stability of asymptotically de Sitter black holes was proved by Mellor and Moss \cite{Mellor-Moss}. This meant that the general relativistic description of black holes is compatible with the idea of the expanding, de Sitter universe. As in four dimensional space-times, there is the uniqueness theorem for Kerr and Reissner-Nordstr\"om solutions, they were generally accepted as most physically relevant.

Last decade, the physical background has considerably changed with appearance of theories, implying existence of extra dimensions in nature, called brane-world theories \cite{ArkaniHamed:1998rs}, \cite{Randall:1999ee}. These theories suggest a solution of the so-called hierarchy problem, that is
the difference in scales of gravitational and electro-weak interactions.
In the scenario with Large Extra Dimensions \cite{ArkaniHamed:1998rs}, the $(3+1)$-dimensional brane, is embedded in a $(4+n)$-dimensional space-time with $n$ space-like compact dimensions. All matter is localized on the $(3+1)$-brane, while fields, which do not carry charge according to the Standard Model gauge group, can propagate in the bulk. An exciting opportunity, that the brane-world theories give, is the possibility to observe the effects of strong, quantum gravity in a laboratory experiment at Tev energies. According to these theories, a miniature black holes might appear in the forthcoming experiments with particle collisions at the Large Hadron Collider or in the Cosmic Showers.

When the black hole radius is much smaller than the characteristic size of extra dimensions, one can describe the black hole by the Schwarzschild-Tangherlini metric \cite{Tangherlini}. When charged particles collide, a charged black holes must be formed. At the same time, recent observational data suggests the non-zero values of the cosmological constant in the Universe, so that the non-vanishing vacuum energy of the world
must influence the formation of black holes. Thus, more general black hole background would be the Reissner-Nordstr\"om-de Sitter (RNdS) generalization of the Schwarzschild-Tangherlini metric. Yet, there is no traditional uniqueness theorem for $D>4$-space-times, so that the important physical criteria that selects from all higher dimensional "black" objects (such as black holes, string, branes, rings, and saturns) is their stability: unstable objects cannot exist or need some mechanism of stabilization.

Nevertheless, the stability analysis of $D \geq 5$ black holes became feasible relatively recently \cite{ishibashi_kodama}, \cite{KodamaKonoplya2}, \cite{Kunduri:2006qa}. This reduction was performed for the D-dimensional Reissner-Nordstr\"om-de Sitter black holes in \cite{ishibashi_kodama} in the general form. Yet, the stability of the Reissner-Nordstr\"om black holes was proven analytically only for $D=4, 5$ space-time dimensions \cite{ishibashi_kodama}. The perturbation equations can be treated separately for all three types, called \emph{scalar}, \emph{vector} and \emph{tensor}, according to the rotation group on the $(D-2)$-sphere. When $D=4$, we know the scalar type as \emph{polar} and the vector type as \emph{axial}, while the tensor type is usually a pure gauge. The higher dimensional cases were addressed in our earlier paper \cite{KonoplyaNPB}, where the stability of the D-dimensional Schwarzschild-de Sitter black holes was proved. Recently the stability of Reissner-Nordstr\"om-anti-de Sitter black holes (without dilaton) was shown in \cite{AdS_stable}. In addition, in \cite{KonoplyaNPB}  the numerical data for the quasinormal modes  for vector and tensor types of gravitational perturbations of Reissner-Nordstr\"om-de Sitter (RNdS) black holes was given. Yet there it was claimed erroneously that Reissner-Nordstr\"om-de Sitter black holes are stable for all values of charge and $\Lambda$-term. In fact, in \cite{KonoplyaNPB} for one particular, and most cumbersome, type of gravitational perturbations, the scalar type, one considered the effective potential, which corresponds to the perturbations of the Einstein equations with the frozen Maxwell field (see Eq. 8 in \cite{KonoplyaNPB}). This approximation is valid when the charge of the black holes $Q$ is considerably less than the black holes mass $M$, yet it is inappropriate for highly charged black holes. In the present paper we consider the dynamic behavior of the wave equation, which corresponds to the complete perturbations of the Einstein-Maxwell equations, given by Eq. (5.61), (5.63 b) in \cite{Kodama:2003kk}.


\begin{figure}
\includegraphics[width=0.75 \linewidth]{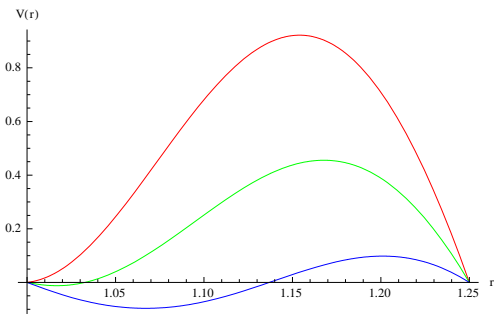}
\caption{The effective potentials $V_-$ for $\rho=0.8$, $q=0.9$, $\ell=2,3,4$ (blue, green red respectively). As $\ell$ grows the peak becomes higher and the negative gap decreases.}\label{potentials_l}
\end{figure}

\textbf{Basic formulae.}  The metric of the $D=d+2$-dimensional RNdS black holes is given by the line element
\begin{equation}\label{metric}
ds^2=f(r)dt^2- f^{-1}(r) dr^2-r^2d\Omega_d.
\end{equation}
where $d\Omega_d$ is the line element on a unit $d$-sphere, $f(r)=1-X+Z-Y,$
\begin{equation}\label{metric-function}
X=\frac{2M}{r^{d-1}},\qquad Y=\frac{2\Lambda r^2}{d(d+1)}, \qquad Z=\frac{Q^2}{r^{2d-2}}.
\end{equation}

The equation of motion for gravitational perturbations of scalar type can be reduced to the wave-like equation
\begin{equation}\label{wave-like}
\left(\frac{\partial^2}{\partial t^2}-\frac{\partial^2}{\partial r_*^2}+V_\pm\right)\Psi(t,r_*)=0,
\end{equation}
where \emph{the tortoise coordinate} $r_*$ is defined as
$dr_*=dr/f(r)$
and the effective potential is a function of black hole parameters, $r$ and of a multipole number $\ell$
that comes from separation of angular variables
\begin{equation}
V_\pm = V_\pm(r, M, Q, \Lambda, D, \ell).
\end{equation}
The explicit form of $V_\pm$ can be found in \cite{Kodama:2003kk}, formulas (5.61-5.63).
The scalar type of gravitational perturbations, corresponding to the $V_{-}$ potential is the only type for which the stability cannot be proved analytically \cite{ishibashi_kodama}, \cite{Kodama:2003kk}. The potential $V_{-}$ reduces to the potential for pure gravitational perturbations, when $Q = 0$. On the contrary,  $V_{+}$ reduces to pure electromagnetic perturbations propagating on the $D$-dimensional Schwarzschild background in the limit of vanishing charge.
%


We shall imply that $\Psi \sim e^{-i \omega t}, \quad \omega = \omega_{Re} - i \omega_{Im},$ so that
$\omega_{Im} > 0$ corresponds to a stable (decayed) mode, while $\omega_{Im} < 0$ corresponds to an unstable (growing) mode.
If the effective potential $V(r)$ is positive definite everywhere outside the black hole event horizon, the differential operator
$d^{2}/dr_{*}^{2} + \omega^{2}$ is a positive self-adjoint operator in the Hilbert space of the
square integrable functions of $r^{*}$, and, any solution of the wave equation with compact support is bounded, what implies
stability. An important feature of the gravitational perturbations is that
the effective potential $V_{-}$ (see Fig. \ref{potentials_l}), which governs the scalar type of the
perturbations, has negative gap for the lower values of the multi-pole numbers $\ell$.
Higher $\ell$ simply increase the top of the potential barrier, and are usually more stable \cite{foot1}. Thus, we shall check here those values of $\ell$, for which the negative gap is present, and therefore the stability is not guaranteed.

\textbf{Numerical Method}.  We shall study the evolution of the black hole perturbations of scalar ``-'' type
in time domain using a numerical characteristic integration method \cite{Gundlach:1993tp}, that uses the light-cone variables $u = t - r_\star$ and $v = t + r_\star$. In the characteristic initial value problem, initial data are specified on the two null surfaces $u = u_{0}$ and $v = v_{0}$. The discretization scheme we used, is
\begin{eqnarray}\label{d-uv-eq}
\Psi(N) &=& \Psi(W) + \Psi(E) - \Psi(S) -\\\nonumber&&-\Delta^2\frac{V(W)\Psi(W) + V(E)\Psi(E)}{8} + \mathcal{O}(\Delta^4) \ ,
\end{eqnarray}
where we have used the following definitions for the points: $N =(u + \Delta, v + \Delta)$, $W = (u + \Delta, v)$, $E = (u, v + \Delta)$ and $S = (u,v)$. This method was very well tested for finding accurate values of the damped quasinormal modes
(see for instance \cite{dampedQNMs} and references therein). Recently it has been also adopted for finding \emph{unstable}, growing, quasinormal modes in \cite{Konoplya:2008yy} for black strings, and in \cite{foot1} for Gauss-Bonnet black holes. The agreement between the time domain and accurate Frobenius methods is excellent. To test the reliability of the method, we increased the precision of the whole numerical procedure and decreased the grid of integration: unchanging of the obtained profiles of $\Psi$ signifies that we have reached sufficient accuracy of the computation.

For convenience, we shall measure all quantities in units of the event horizon $r_+$. Since the value of the event horizon is $r_+=1$, the black hole mass is fixed as $2M=1+Q^2- 2 \Lambda/d(d+1)$. The cosmological constant written in terms of the cosmological horizon $r_c$ allows
to introduce the convenient variable $\rho=r_+/r_c=1/r_c<1$, so that
\begin{equation}
\Lambda=\rho^2\frac{d(d+1)}{2}\frac{(1+Q^2)(\rho^{d-1}-1)}{\rho^{d+1}-1}\,.
\end{equation}
We shall consider also the charge normalized by its extremal quantity $q=Q/Q_{ext}<1$.

\begin{figure}
\includegraphics[width=0.75 \linewidth]{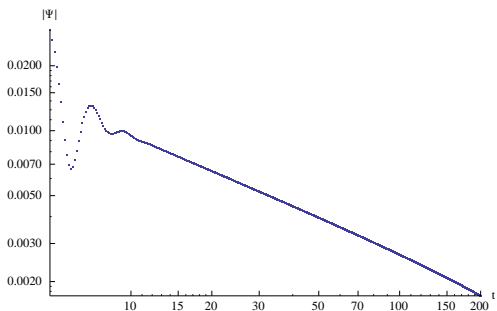}
\caption{Time-domain profile of near extremal $q=0.999$ Reissner-Nordstr\"om black hole perturbation ($D=11$, $\rho=0$). At the late time the power-law tail is observed (straight line in the logarithmic scale). The epoch of the quasinormal oscillations becomes shorter for near extremal $Q$. }\label{power-tail}
\end{figure}


\begin{figure}
\includegraphics[width=0.75 \linewidth]{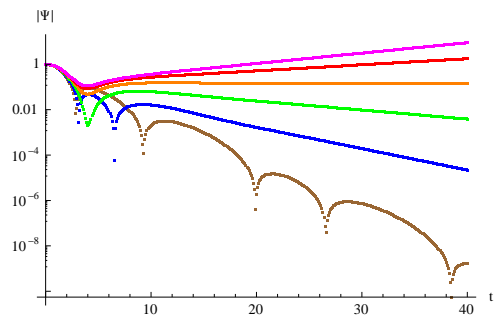}
\caption{Time-domain profile of near extremal Reissner-Nordstr\"om black hole perturbation ($D=11$, $\rho=0.8$).
q=0.4 (brown) q=0.5 (blue) q=0.6 (green) q=0.7 (orange) q=0.8 (red)
q=0.9 (magenta). The smaller q, the slower growth of the profile is.}\label{RNdS.profiles.D=11.rho=0.8}
\end{figure}

\begin{figure}
\includegraphics[width=0.75 \linewidth]{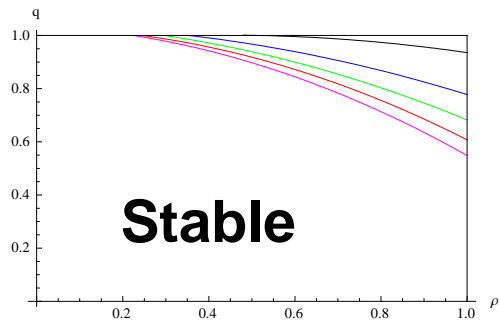}
\caption{The parametric region of instability in the right upper corner of the square in the $\rho-q$ "coordinates" for $D=7$ (top, black), $D=8$ (blue), $D=9$ (green), $D=10$ (red), $D=11$ (bottom, magenta).}\label{parametrs_instability}
\end{figure}

\begin{table*}
\caption{$l=2$ fundamental frequencies of gravitational perturbations of scalar type $V_-$ of $D$-dimensional RNdS black holes.}
\begin{tabular}{|l|c|c|c|c|c|c|c|}
\hline
$q$&$D=5$&$D=6$&$D=7$&$D=8$&$D=9$&$D=10$&$D=11$\\
\hline
$0$&$0.948-0.256\imo$&$1.137-0.304\imo$&$1.339-0.401\imo$&$1.564-0.603\imo$&$1.997-0.863\imo$&$2.460-0.987\imo$&$2.902-1.087\imo$\\
$0.1$&$0.941-0.254\imo$&$1.130-0.302\imo$&$1.332-0.401\imo$&$1.558-0.608\imo$&$1.998-0.862\imo$&$2.459-0.983\imo$&$2.900-1.083\imo$\\
$0.2$&$0.922-0.247\imo$&$1.110-0.296\imo$&$1.311-0.400\imo$&$1.545-0.623\imo$&$2.001-0.856\imo$&$2.456-0.973\imo$&$2.895-1.072\imo$\\
$0.3$&$0.894-0.237\imo$&$1.080-0.289\imo$&$1.282-0.402\imo$&$1.537-0.646\imo$&$2.002-0.844\imo$&$2.449-0.956\imo$&$2.884-1.053\imo$\\
$0.4$&$0.859-0.225\imo$&$1.045-0.282\imo$&$1.248-0.406\imo$&$1.545-0.660\imo$&$1.997-0.822\imo$&$2.435-0.932\imo$&$2.866-1.029\imo$\\
$0.5$&$0.821-0.213\imo$&$1.007-0.276\imo$&$1.219-0.414\imo$&$1.552-0.649\imo$&$1.984-0.794\imo$&$2.412-0.903\imo$&$2.840-1.001\imo$\\
$0.6$&$0.782-0.201\imo$&$0.970-0.271\imo$&$1.198-0.419\imo$&$1.545-0.624\imo$&$1.959-0.764\imo$&$2.380-0.875\imo$&$2.805-0.974\imo$\\
$0.7$&$0.742-0.190\imo$&$0.938-0.267\imo$&$1.180-0.412\imo$&$1.522-0.596\imo$&$1.925-0.737\imo$&$2.342-0.851\imo$&$2.764-0.953\imo$\\
$0.8$&$0.705-0.181\imo$&$0.908-0.260\imo$&$1.156-0.399\imo$&$1.490-0.575\imo$&$1.888-0.720\imo$&$2.303-0.836\imo$&$2.725-0.938\imo$\\
$0.9$&$0.670-0.172\imo$&$0.878-0.252\imo$&$1.128-0.387\imo$&$1.459-0.562\imo$&$1.855-0.707\imo$&$2.268-0.823\imo$&$2.689-0.926\imo$\\
$0.98$&$0.643-0.165\imo$&$0.854-0.245\imo$&$1.107-0.380\imo$&$1.435-0.552\imo$&$-$&$-$&$-$\\
\hline
\end{tabular}
\end{table*}

\textbf{Discussion of the results}. Let us start from the pure Reissner-Nordstr\"om (RN) black holes ($\rho =0$). From the Table I
one can see that quasinormal modes of non-extremal pure RN holes are damped for $D=6, 7, ..11$.
For the near extremal values of charge $Q$, a power-law damped tail dominates at asymptotically late times (Fig. \ref{power-tail}).
When approaching near extremal $Q$, the epoch of quasinormal oscillations becomes much shorter (Fig. \ref{power-tail}), so that it is difficult to find out the accurate values of the QN frequency from the time domain profile, especially for higher $D$. Therefore some values in Table I are absent.

The RNdS black holes are characterized by non-zero values of $\rho$ and $q$.
After careful testing of all the range of values of parameters $q$ and $\rho$, we have found that for sufficiently large values of
the \emph{both} parameters, $D \geq 7$ RNdS black holes are unstable. The typical picture of developing of instability can be found on
Fig. \ref{RNdS.profiles.D=11.rho=0.8}. There one can see that for moderate values of $q$, the profile consists of damped quasinormal oscillations.
Then, as the charge increases, the real oscillation frequency of the ringing decreases, approaching zero in the threshold point of instability.
This is well understood, because unstable modes must be pure imaginary and the threshold point of instability corresponds to some static solution $\omega = 0$ of the wave equation \cite{Konoplya:2008yy}.  This is a natural picture for instability developed at lowest multi-poles.
Instability induced by large $\ell$, on the contrary, appears as the growing of $\Psi$  after a long period of damped oscillations
\cite{foot1}.

The parametric region of instability is shown on Fig. \ref{parametrs_instability}. For larger $D$ the region of instability is bigger.
Another interesting question, which was beyond the scope of our paper, is if the extremal $D=6$ Reissner-Nordstr\"om-de Sitter black holes are stable? Within the numerical method we can approach quite near the extremal values, but not the exact extremal limit. 


\textbf{Black hole deformation at the edge of stability}  Let us study in detail situation when the charge is near the threshold of stability. We have found that there is \emph{a static mode} ($\omega=0$) for $\ell=2$ ``-'' type of perturbations, while all other oscillations decay. We shall find here the form of the perturbed metric at the threshold point of instability in the linear approximation as a background metric plus small perturbation, which is approximated by the found static mode. The correction to the black hole metric is given by Eq. (5.6a) of \cite{Kodama:2003kk}
$$ \delta g_{ab} = f_{ab}(r){\mathbb S}(z), \quad \delta g_{ai} = r f_a(r){\mathbb S}_i(z), $$
\begin{equation}
\delta g_{ij} = 2r^2(H_L(r)\gamma_{ij}{\mathbb S}(z) + H_T(r) {\mathbb S}_{ij}(z)),
\end{equation}
where ${\mathbb S}(z)$ are the scalar harmonics, which satisfy
\begin{equation}
{\hat\Delta}{\mathbb S}(z)=\ell(\ell+d-1){\mathbb S}(z).
\end{equation}
A unit $d$-sphere is defined by the metric
$$d\Omega_d^2=\gamma_{ij}(z)dz^idz^j,$$
with the angular variables $\{z\}$.

From the Eq. (5.37) \cite{Kodama:2003kk} we see that $\tilde F^r_t=0$ when $\omega=0$. Consequently, in the Regge-Willer gauge ($f_a(r)=0$, $H_T(r)=0$) we can see that the perturbed metric is static because $f_{01}(r)=0$ (see Eqs. (5.7b) and (5.8) in \cite{Kodama:2003kk}). Thus, the event horizon surface is given by the equation
\begin{equation}
g_{00}(r)+f_{00}(r){\mathbb S}(z)=0.
\end{equation}
Expanding this equation near the unperturbed black hole horizon $r_+$ we find the expression for the black hole horizon $\tilde r_+$ of the perturbed black hole
\begin{equation}\label{newhor}
\tilde r_+\approx r_+-\frac{f_{00}(r_+){\mathbb S}(z)}{g_{00}'(r_+)}=r_++\frac{f_{00}(r_+){\mathbb S}(z)}{f'(r_+)}.
\end{equation}

The radial coordinate $R$ of the perturbed black hole is related to the coordinate $r$ as
\begin{equation}\label{newradial}
R=r\sqrt{1+2H_L(r){\mathbb S}(z)}\approx r(1+H_L(r){\mathbb S}(z)).
\end{equation}
Substituting (\ref{newhor}) into (\ref{newradial}) we find for the perturbed horizon
$R_+=r_++{\cal A}{\mathbb S}(z)$, where ${\cal A}$ is proportional to the amplitude of the linear perturbation
$${\cal A} \approx f_{00}(r_+)/f'(r_+) + r_+H_L(r_+).$$

\begin{figure}
\includegraphics[width=0.5\textwidth]{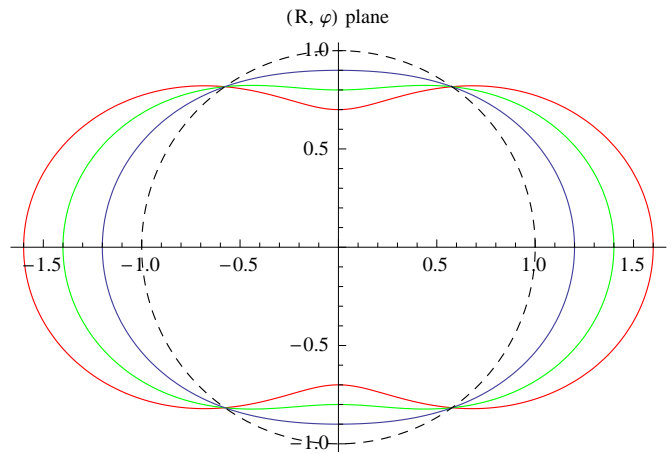}
\caption{\label{BH_perturbed} The equatorial plane of the black hole horizon hypersurfaces at the edge of stability. The dashed line corresponds to the unperturbed black hole of unit horizon. The blue, green and red lines correspond to the perturbed black holes after decay of all the dynamical modes.}
\end{figure}

We see that the null hypersurface is the sphere corrected by a small perturbation proportional to ${\mathbb S}_{\ell=2}(z)$. This surface can be easily imagined, because we always can choose such an angular coordinate $\phi$ (associated with the largest azimuthal number $m=2$) that ${\mathbb S}_{\ell=2}(z)={\mathbb S}_{\ell=2}(\phi)$ and ${\mathbb S}_{\ell=2}(\phi)\propto cos(2\phi)+const$.
Figure \ref{BH_perturbed} of the shape of the perturbed black hole at the instability suggests that the black hole could either split into two black holes or to transform into another solution, such as, for instance, a black ring. The splitting into two black holes looks nevertheless improbable, because such smaller holes would be themselves unstable and overcharged.

One should note that since we consider a linear approximation, we cannot describe the instability growth at later times because initially small perturbations will become comparable with the background at sufficiently late times. The final stage of the evolution of a unstable state must be described by the fully non-linear theory. The linear theory provides us only an initial guess for the developing of instability.

\textbf{Conclusions} Let us enumerate the obtained results.

1) The Reissner-Nordstr\"om black holes are stable for $D = 5, 6,..11$.

2) The $D \geq 7$ Reissner-Nordstr\"om-de Sitter black holes are unstable if values of the black hole charge and mass are large enough.

3) The threshold values of parameters $q$ and $\rho$, for which the instability appears, correspond to the dominance of some static solution of the wave equation.

4) The shape of the slightly perturbed black hole at the instability point has been found.

Strictly speaking the above instability does not mean that the corresponding black holes cannot exist, for instance
a universe with considerable values of the cosmological constant (presumably our universe in the early epochs), should have the chromo-dynamics instead of the considered here U(1) electrodynamics. If nevertheless one wants to preserve $U(1)$ electrodynamic, higher dimensional asymptotically de Sitter world, and large number of extra-dimensions $D \geq 7$ at the same time, he should take into account the above instability of black holes. This is important when investigating quasinormal modes or Hawking radiation of the Standard Model fields on the higher dimensional Reissner-Nordstr\"om-de Sitter background \cite{QNM_Haw}. The absence of coincidence of gravitational and thermodynamic instabilities is also easily understood, if one notices that the thermodynamic instability is induced by quantum effects, while gravitational instability is a classical effect.


\textbf{Acknowledgments} A.Z. was supported by \emph{Funda\c{c}\~ao de Amparo \`a Pesquisa do Estado de S\~ao Paulo (FAPESP)}, Brazil.
R.A.K. was supported by \emph{the Japan Society for the Promotion of Science}. R. A. K. acknowledges interesting discussions with
H.Reall and G.Gibbons.

\end{document}